\documentclass[%
reprint,
prb,
superscriptaddress,
 amsmath,amssymb,
 aps,
floatfix,
]{revtex4-2}

\usepackage{fix-cm}
\usepackage{graphicx}
\usepackage{dcolumn}
\usepackage[colorlinks=true,linkcolor=blue]{hyperref}

\usepackage{bm}
\usepackage{amsmath}
\usepackage{amsfonts}
\usepackage{amssymb}

\usepackage[colorlinks=true,linkcolor=blue]{hyperref}

\usepackage[utf8]{inputenc}
\usepackage[T1]{fontenc}
\usepackage{mathptmx}
\usepackage{etoolbox}

\begin{document}

\preprint{APS/123-QED}

\title{Magnetic texture control in ion-implanted metamaterials}

\author{Christina Vantaraki}
\email[Contact author: ]{christina.vantaraki@physics.uu.se}
\affiliation{Department of Physics and Astronomy, Uppsala University, Box 516, 75120 Uppsala, Sweden}

\author{Kristina Ignatova}
\affiliation{Science Institute, University of Iceland, Dunhaga 3, Reykjavik 107, Iceland}

\author{Dmitrii Moldarev}
\affiliation{Department of Physics and Astronomy, Uppsala University, Box 516, 75120 Uppsala, Sweden}

\author{Matías P. Grassi}
\affiliation{Department of Physics and Astronomy, Uppsala University, Box 516, 75120 Uppsala, Sweden}

\author{Michael Foerster}
\affiliation{ALBA Synchrotron Light Facility, 08290-Cerdanyola del Valles, Barcelona, Spain}

\author{Daniel Primetzhofer}
\affiliation{Department of Physics and Astronomy, Uppsala University, Box 516, 75120 Uppsala, Sweden}

\author{Unnar B. Arnalds}
\affiliation{Science Institute, University of Iceland, Dunhaga 3, Reykjavik 107, Iceland}

\author{Vassilios Kapaklis}
\email[Contact author: ]{vassilios.kapaklis@physics.uu.se}
\affiliation{Department of Physics and Astronomy, Uppsala University, Box 516, 75120 Uppsala, Sweden}

\begin{abstract}
We study experimentally the impact of the additive fabrication method on the magnetic properties of Fe$^+$-implanted Pd square artificial spin ice lattices. Our findings show that the lattices exhibit a higher ordering temperature than their continuous film counterparts.
This behavior is attributed to the additive fabrication process, which induces an inhomogeneous Fe concentration within the lattice building blocks. 
Moreover, the implantation process creates a magnetic depth profile, enabling temperature-dependent tunability of the magnetic thickness. These additional internal degrees of freedom broaden the design possibilities for magnetic metamaterials, allowing precise fine-tuning of their static and dynamic properties to achieve complex and customizable behaviors.

\end{abstract}

\maketitle

\section{\label{Intro}Introduction}
Complex systems are characterized by interactions among their components that lead to the spontaneous emergence of collective properties. In condensed matter physics, magnetic metamaterials have become an exemplary platform for investigating such complexity. These systems consist of arrays of ferromagnetic elements of sub-$\mu$m size that interact magnetostatically \cite{HeydermanLauraJ.2021MmsF,HeydermanLJ2013Afsn}, with these interactions giving rise to emergent phenomena, including collective magnetic order \cite{Wang:2006kt,StopfelHenry2018Moae,OstmanErik2018Imia,PerrinYann2016EdCp,QiYi2008Doot,BranfordW.R2010Doom} and dynamics \cite{KapaklisVassilios2012Masi,MarrowsChristopherH2011Tgoa,MorleyS.A.2017Vfoa,MorleySophieA2015Tamd,GligaSebastian2020Dora,Kapaklis:2014ea,AnderssonM.S.2016Timr}. The building blocks of magnetic metamaterials, mesoscopic ferromagnetic elements, are often approximated as point-like magnetic dipoles or artificial magnetic atoms. While these analogies are useful, they fall short in capturing phenomena such as thermal fluctuations and phase transitions. Recent studies have provided a more nuanced analysis of the intrinsic magnetic properties and dynamics of these elements \cite{Gliga_edge_entropy_2015, Sloetjes_APL, Sloetjes_PRB, Sloetjes_JPCM, Sloetjes_PRM}. In a notable example, the initially simulated internal degrees of freedom, leading to excess entropy and modification of dynamics as reported by \citet{Gliga_edge_entropy_2015}, were later experimentally verified, with their temperature-dependent behavior characterized in detail by \citet{SkovdalBjörnErik2023Tewa}. These systems bridge theoretical models of complexity with experimental realizations, enabling studies of the interplay between competing interactions \cite{NguyenV.-D.2017Ciia,vantaraki2024magneticorderlongrangeinteractions,LouHaifeng2023Ciip}, frustration, and exotic magnetic phases \cite{Gilbert:2016cn, Nisoli:2017hg}. In addition to their fundamental significance, the complexity in magnetic metamaterials has fueled technological advancements in fields such as spintronics \cite{ASI_Review_PSI}, data storage \cite{Gartside:2017kc}, and unconventional computation \cite{GartsideJackC.2020Cnwf, Task-adaptive_RC_2024}.

In this work, we explore how magnetic metamaterials can serve as a platform to deepen our understanding of complexity in physics. Using electron-beam lithography and ion implantation techniques \cite{vantaraki2024magneticmetamaterialsionimplantation}, we fabricated ferromagnetic elements with intrinsic compositional and magnetic inhomogeneities. Our investigation examines the impact of these inhomogeneities on the ordering temperature of extended arrays and the internal magnetic textures of individual elements. The study focuses on square artificial spin ice (ASI) lattices composed of elongated, stadium-shaped nanoelements—referred to as \textit{mesospins}. While ASIs have been widely studied for emergent order \cite{Wang:2006kt,OstmanErik2018Imia} and thermally induced dynamics \cite{KapaklisVassilios2012Masi, Pohlit:2020hv, SkovdalBjörnErik2023Tewa} in mesospins made from isotropic and homogeneous magnetic materials, our work provides a direct comparison with systems exhibiting intrinsic mesospin inhomogeneities. This comparison not only enhances our understanding of their fundamental behavior but also underscores their potential for practical applications \cite{Gliga_edge_entropy_2015}, as it enables the deliberate design of both the static and dynamic properties of magnetic metamaterials.

\section{\label{Materials_methods}Materials and Methods}

\subsection{Sample fabrication}
\begin{figure}[ht]
\includegraphics[width=0.95\linewidth]{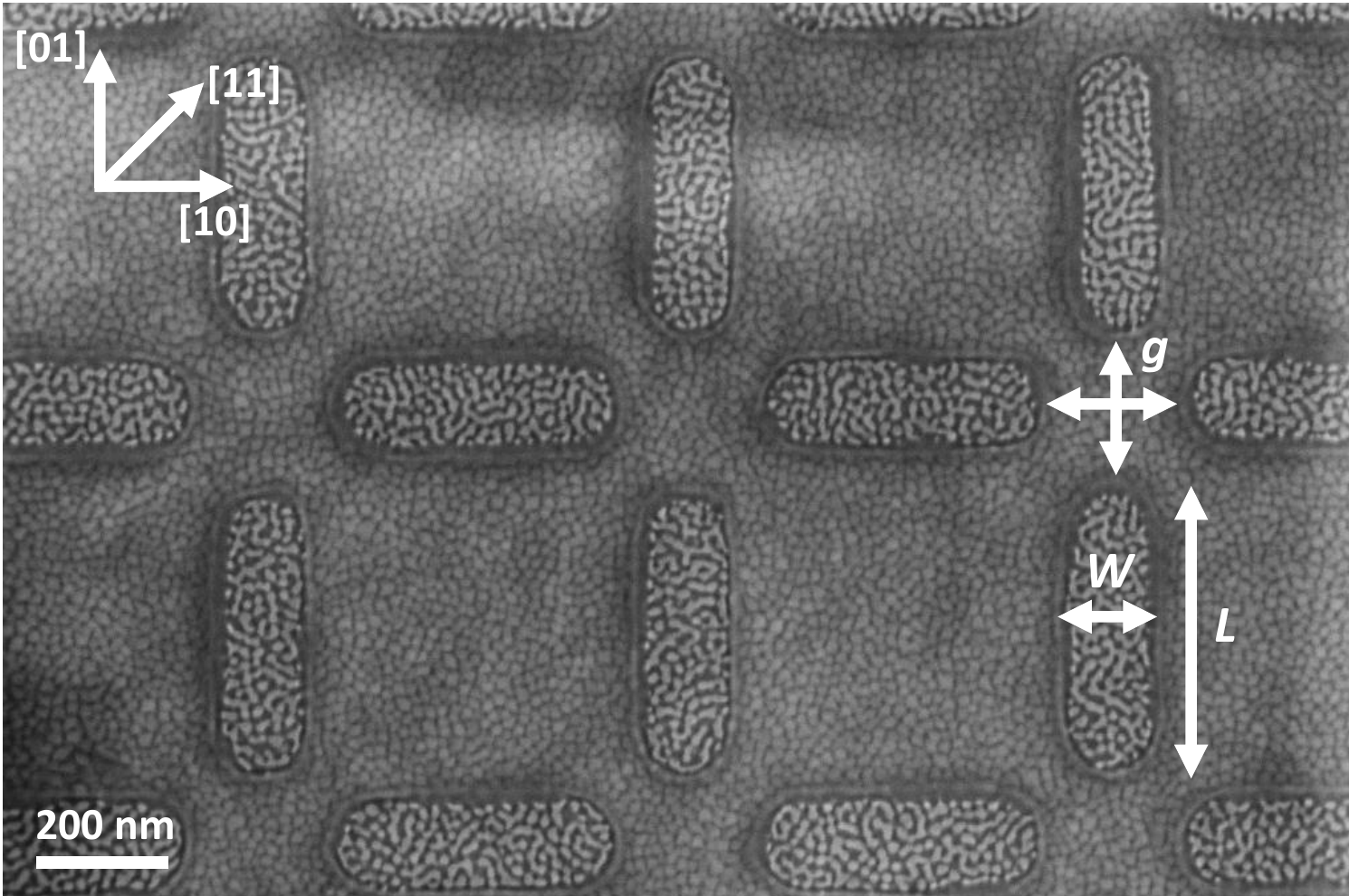}
\caption{Scanning electron microscopy (SEM) image of implanted square artificial spin ice lattice (ASI) with nominal fluence 2.7$\times$10$^{16}$~ions/cm$^{2}$.}
\label{Figure1}
\end{figure}

Two samples were studied, having different Fe concentrations. Both were fabricated by implanting 30 keV $^{56}$Fe$^{+}$ into a Pd film through a patterned Cr mask. The Pd film, deposited using DC magnetron sputtering on a MgO substrate, consisted of a 60 nm Pd layer with a 5 nm V adhesion layer and a 6 nm Cr capping layer. This additive fabrication process yields flat Fe$_{x}$Pd$_{100-x}$ (where $x$ stands for at.\%) ferromagnetic structures, where Fe has a concentration-depth profile. A detailed description about the fabrication process is provided in \citet{vantaraki2024magneticmetamaterialsionimplantation}. One sample was implanted with a nominal fluence of 2.7$\times$10$^{16}$ ions/cm$^{2}$, while the other was implanted with 3$\times$10$^{16}$ ions/cm$^{2}$. 
Higher fluences yield higher Fe concentration \cite{StromPetter2022Soft}. In both samples, the stadium-shaped elements have a length of $L$~=~470 nm and width $W$~=~170 nm, while the edge-to-edge gap between the elements is $g$~=~170 nm (see Fig. \ref{Figure1}). 

\subsection{Monte-Carlo simulations}
The Fe distribution within the implanted samples was investigated by performing Stopping and Range of Ions in Matter (SRIM) simulations \cite{ZieglerJamesF.2010S–Ts}. To imitate the experimental conditions, the simulations were performed for 30~keV Fe$^{+}$ ions implanted in a 600~Å Pd film through a 60~Å Cr capping layer. The angle of ion incidence was set to zero degrees (normal to the surface). 

\subsection{Hysteresis protocols}
The temperature-dependent hysteretic behavior of implanted ASIs was studied using a longitudinal magneto-optical Kerr effect (L-MOKE) setup and a vibrating sample magnetometry (VSM) system. 
In the L-MOKE setup, a $p$-polarized incident laser beam with a wavelength of 659 nm was modulated by a Faraday cell and subsequently reflected by the sample. The reflected beam was passed through an analyzer (with extinction ratio of 10$^5$) and then measured using a biased Si photodiode. The detector was connected with a pre-amplifier and a lock-in amplifier. 
For a single hysteresis loop measurement, the external magnetic field was cycled from +$H_{max}$ to -$H_{max}$ and then back to +$H_{max}$, where $H_{max}$ = 50~mT. To increase the signal-to-noise ratio, 50 hysteresis loops were recorded at each temperature, and the final hysteresis loop was obtained by averaging these measurements. The external magnetic field was applied as a sinusoidal waveform $H = H_{0} sin(2 \pi ft)$, with $f$ = 0.1~Hz. The field was applied along the [11] and [01] directions of the ASI lattice, as indicated in Fig.~\ref{Figure1}.
The 659 nm laser beam achieves a penetration depth of approximately 12 nm in the Pd film (see Supplemental Material), detecting the magnetization response from the surface of the sample. The cryostat in the L-MOKE system operates up to 400~K.

In addition to MOKE measurements, magnetic hysteresis loops were recorded by Vibrating Sample Magnetometry (VSM) with a 2~T Lakeshore 8600 magnetometer, equipped with a single-stage variable temperature unit that can reach a temperature of 950~K. This technique records the magnetic response from the entire sample volume.

\subsection{Magnetic imaging}
The magnetic state of the implanted mesospins was directly imaged using photoemission electron microscopy, employing x-ray magnetic circular dichroism (PEEM-XMCD). These experiments were conducted at the CIRCE (bl24) beamline of the ALBA synchrotron \cite{AballeLucia2015TAsL}. 
The images were acquired at room temperature with an x-ray photon energy tuned to the Fe L$_3$-edge. The lattices were imaged at the remanent state after applying a high enough field to saturate the array \cite{FoersterMichael2016Csea}.

\subsection{Micromagnetic simulations} \label{sectionB}
The magnetic texture of the inhomogeneous mesospins was explored by performing micromagnetic simulations using MuMax$^{3}$ \cite{VansteenkisteArne2014Tdav}. The simulations were performed with stadium-shaped elements of dimensions 470 $\times$ 170 nm$^{2}$ arranged on a square lattice. The edge-to-edge gap between the elements was $g$ = 170~nm. To mimic the experimental conditions, the elements were considered to be composed of Fe with a thickness of 3.18~nm. This thickness corresponds to a continuous Fe layer implanted with a nominal fluence of 2.7$\times$10$^{16}$~ions/cm$^{2}$ (see Supplemental Material). It is important to note that this is a rough approximation, as the fabrication process yields Fe$_{x}$Pd$_{100-x}$ (where $x$ stands for at.\%) elements. Consequently, the values for saturation magnetization and exchange stiffness constant in simulations deviate from those of the actual material. The elements were discretized into grid cells with cell size smaller than the exchange length and equal to 2.5~nm. 

\section{Results and Discussion}
\begin{figure}[t]
\includegraphics[width=1\linewidth]{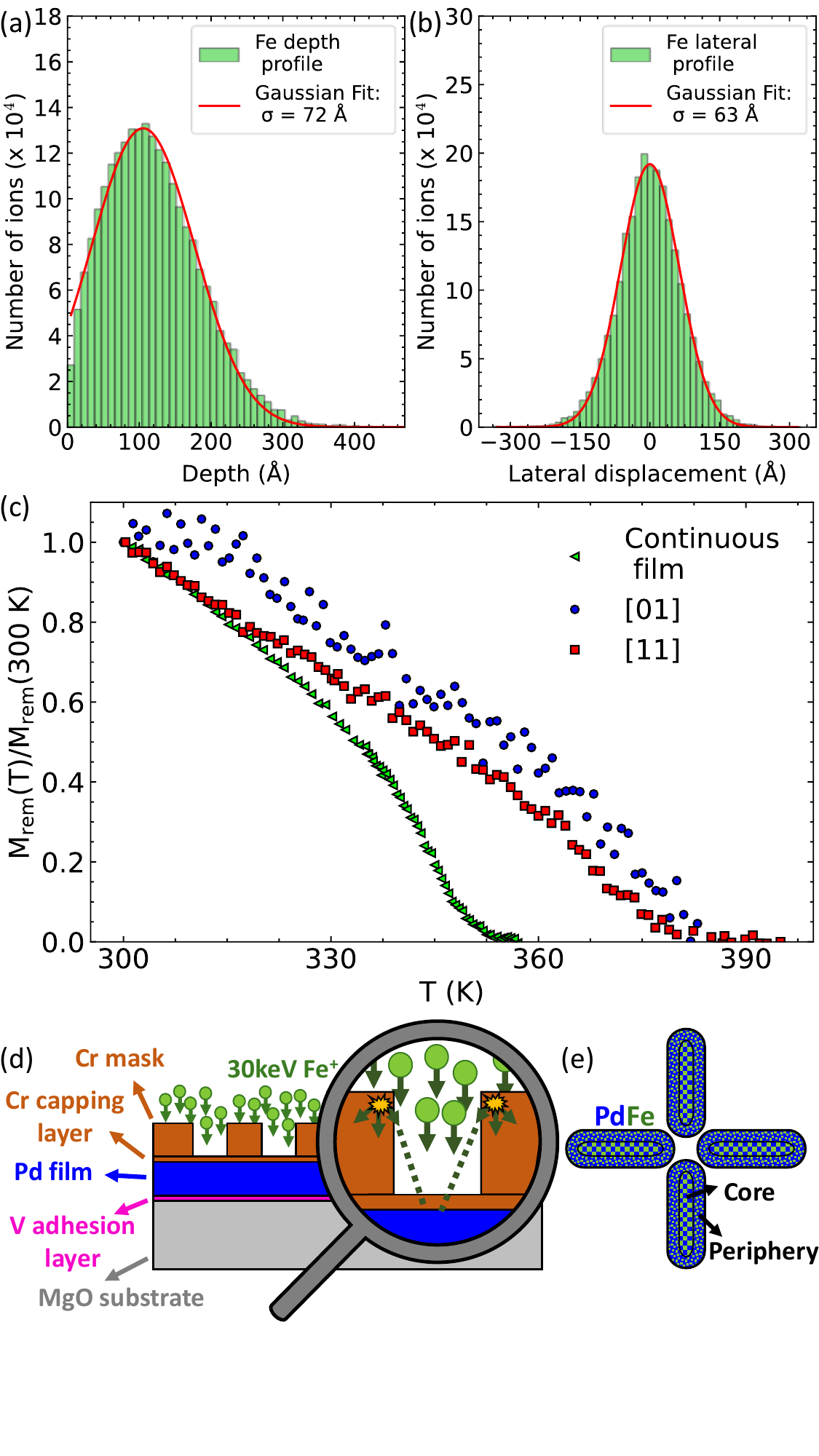}
\caption{(a,b) The simulated (a) depth- and (b) lateral-profile of Fe as a consequence of the ions straggling along the beam and perpendicular to it. (c) Normalized remanent magnetization M$_{\textrm{rem}}$ as a function of temperature T for the ASI and its reference continuous film implanted with 30~keV Fe of nominal fluence 2.7$\times$10$^{16}$~ions/cm$^{2}$ measured with L-MOKE. (d) Schematic representation of the side view of the patterned film and the scattering events during Fe implantation. Because of the presence of the mask, the implanted mesospins have higher Fe concentration compared to the reference continuous film. (e) The Fe concentration gradually decreases toward the edges of the elements, resulting in a higher concentration at their cores compared to the peripheries.} 
\label{Figure2}
\end{figure}

We begin by examining the distribution of the implanted Fe on the Pd film. To do so, we perform Monte Carlo simulations using the SRIM software. The simulations provide information about the Fe$^{+}$ distribution both in depth and laterally from single implantation point, assuming an amorphous Pd target.
As shown in Fig. \ref{Figure2} (a) and (b), the simulation results show that implanted Fe exhibits a concentration depth-profile [Fig. \ref{Figure2} (a)] and a concentration lateral-profile [Fig. \ref{Figure2} (b)]. Both depth- and lateral-profile follow approximately a Gaussian distribution. In the concentration depth-profile, the peak of the distribution corresponds to the average depth at which the ions come to rest, while in the concentration lateral-profile, the peak indicates the surface spot that receives the highest ion fluence. For both profiles, the standard deviation $\sigma$ of the Gaussian distribution represents the extent of the ion spread around the peak, indicating the depth variation ($\sigma$ = 72~Å) in the depth-profile and the lateral spread or straggling ($\sigma$ = 63~Å) in the lateral-profile. The concentration depth-profile has been investigated previously in continuous films \cite{vantaraki2024magneticmetamaterialsionimplantation, StromPetter2022Soft}. In contrast, the lateral implantation profile is often neglected in continuous films, as the lateral motion of ions creates a broader distribution due to the system's extension. However, the concentration lateral-profile can no longer be neglected in the lattices due to the clear boundary between the implanted and unimplanted regions \cite{FAYEMM1992Tddi}.

The concentration profiles are difficult to be probed in lattices with nanometer elements; however, the amount of implanted Fe can be investigated using the ordering temperature as the probe.
The ordering temperature is defined as the temperature at which the spontaneous magnetization vanishes. To a good approximation, this temperature coincides with the temperature where the remanent magnetization disappears \cite{KERKMANND1992TmaC}. 
As such, we examine the temperature-dependent behavior of the remanent magnetization M$_{\textrm{rem}}$ from the MOKE measurements for an implanted ASI array and its reference continuous film that was prepared and implanted next to the lattice, with a nominal fluence of 2.7$\times$10$^{16}$~ions/cm$^{2}$.
As shown in Fig. \ref{Figure2}~(c), the remanent magnetization vanishes at higher temperatures for the implanted ASI compared to the reference implanted film, reflecting a higher ordering temperature. This trend is opposite to the one previously observed in conventional ASI lattices \cite{KapaklisVassilios2012Masi}. It is worth noting that the collapse of the magnetic order occurs at a lower temperature for the continuous film compared to the lattice even when considering the saturation magnetization (see Supplemental Material).
This observation highlights a higher Fe concentration for the patterned film.
As shown schematically in Fig. \ref{Figure2}~(d), the additional Fe concentration in ASI is attributed to Fe ions that impinge the mask near the edges. Due to multiple scattering events, some ions manage to scatter back out of the mask and then become embedded in the Pd film at an angle. 
As a consequence, the elements are enriched with additional Fe dopants, which are not implanted in the continuous reference film, because of the presence of the mask. Based on the increase in critical temperature, we estimate that the ASI lattice has approximately 1-3\% higher Fe concentration compared to the continuous film. This estimation was derived from the relationship between the Fe concentration and the ordering temperature found in Refs.  \cite{HITZFELDM1984FoPa,StromPetter2022Soft,RohmanL2019Ctac}.
Nevertheless, the additional dopants cannot prevent the lateral spread of ions at the edges of the elements. 
As a consequence, the Fe concentration gradually decreases toward the edges of the elements \cite{StromPetter2024PIoF}. This effect results in a higher Fe concentration in the core area of the elements compared to their periphery [Fig.~\ref{Figure2}~(e)].
Similar mask proximity effects have been reported for CMOS devices \cite{HookT.B.2003Liis,HoblerG.1989MCso,Yi-MingSheu2005Mwep}.

\begin{figure}[t!]
\includegraphics[width=1\linewidth]{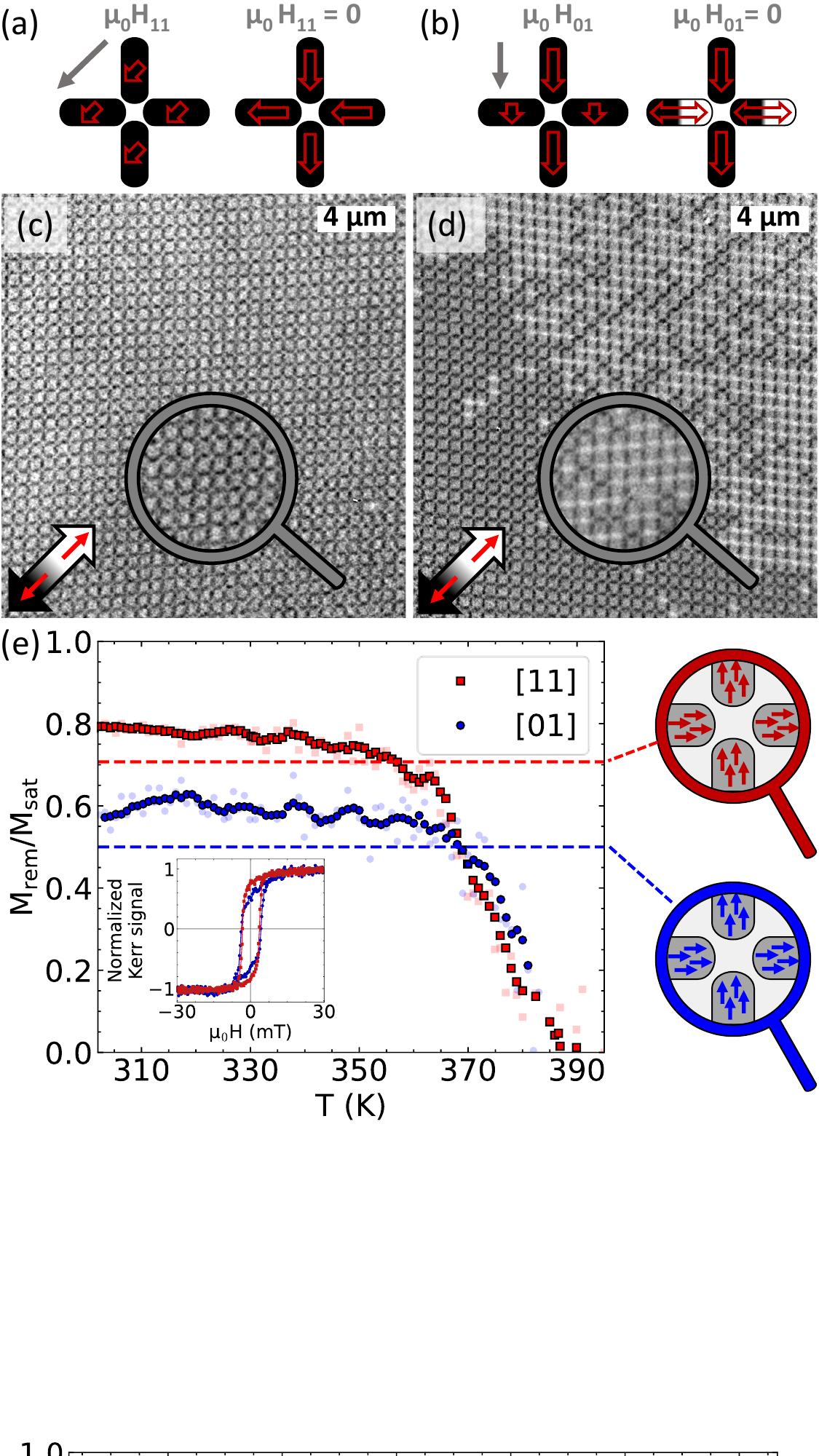}
\caption{(a,b) Schematic representation of the magnetization orientation for an ASI array at the saturation (left) and remanence state (right) for a field applied along the (a) [11] and (b) [01] direction. (c,d) PEEM-XMCD images for implanted ASI lattices of nominal fluence 2.7$\times$10$^{16}$~ions/cm$^{2}$. The images were captured at the remanence state after the field was applied along the (c) [11] and (d) [01] direction. The PEEM-XMCD images are distorted due to electron microscope optics artifacts. (e) The ratio between the remanence and saturation magnetization, M$_{\textrm{rem}}$/M$_{\textrm{sat}}$, extracted from the L-MOKE measurements for the implanted ASI when the field was applied along the [11] and [01] direction. The saturation magnetization was determined from the hysteresis loops by averaging the values between 30~mT and 50~mT. Both the raw data (transparent points) and the smoothed data (non-transparent points), obtained using a five-point running average, are presented. The inset shows representative hysteresis loops for the ASI lattice. The dotted lines indicate the M$_{\textrm{rem}}$/M$_{\textrm{sat}}$ values for collinear magnetic texture for the [11] (red) and [01] (blue) direction.}
\label{Figure3}
\end{figure}

To directly illustrate whether the inhomogeneities influence the magnetic state of mesospins, we image the ASI array in the remanent state by performing XMCD-PEEM experiments, 
after applying an external in-plane field to saturate the array. The field was applied along the diagonal of the elements, corresponding to the [11] direction [Fig.~\ref{Figure3} (a), left], and parallel to the long-axis of half of the elements, corresponding to the [01] direction [Fig.~\ref{Figure3} (b), left]. 
Representative XMCD-PEEM images are shown in Fig. \ref{Figure3} (c) and (d), when the field was applied along the [11] and [01] direction, respectively. In both images, the elements are displayed in a uniform white or black contrast. The white and black colors indicate a magnetization component parallel and antiparallel to the X-ray beam, respectively. Consequently, we conclude that
the ASI array has obtained a ferromagnetic (or Type II) remanent state when the field is applied along the [11] direction. Similarly, applying a field along the [01] direction and then returning to the remanent state, results in a full ferromagnetic order for the elements parallel to the [01] direction. For the mesospins parallel to the [10] direction, a mixture of ferromagnetic and antiferromagnetic configurations is observed. However, the ferromagnetic order is more dominant than the antiferromagnetic one, as it creates vertices of lower energy (E$_\mathrm{{Type \: II}}$ < E$_\mathrm{{Type \: III}}$) (see Supplemental Material). These observed remanent states confirm that shape anisotropy determines the magnetization direction, as expected [Fig.~\ref{Figure3}~(a) and (b)], while they indicate a dominant collinear magnetization component within the elements. However, the resolution of the XMCD-PEEM technique does not allow the detection of magnetization components not aligned with the X-ray beam, thereby preventing the investigation of non-collinearities in the magnetization texture.

\begin{figure}[t]
\includegraphics[width=1\linewidth]{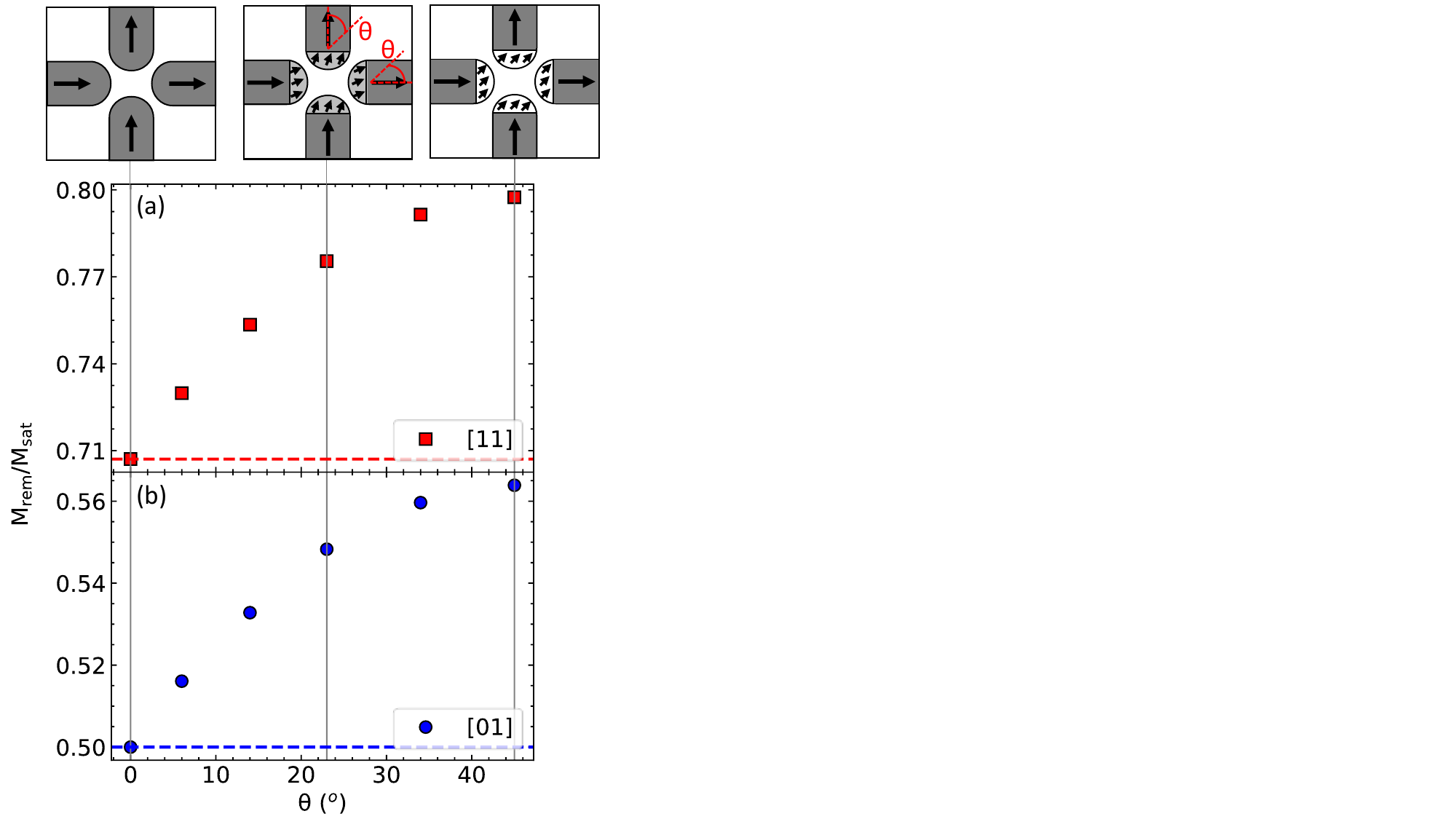}
\caption{The simulated ratio M$_{\textrm{rem}}$/M$_{\textrm{sat}}$ as a function of the angle $\theta$ between the collinear and non-collinear magnetization components if a field had been applied along the (a) [11] and (b) [01] direction. The dotted lines indicate the M$_{\textrm{rem}}$/M$_{\textrm{sat}}$ values for collinear magnetic texture.}
\label{Figure4}
\end{figure}

To investigate the presence of magnetization bending, we examine the remanent magnetization normalized by the saturation magnetization, M$_{\mathrm{rem}}$/M$_{\mathrm{sat}}$, obtained from the L-MOKE measurements. This quantity is an indirect way of determining if the magnetization texture within the elements is fully collinear \cite{SkovdalBjörnErik2023Tewa}. For the observed remanent states of an ASI lattice with fully collinear magnetization, M$_{\mathrm{rem}}$/M$_{\mathrm{sat}}$ has been previously shown to take characteristic values: 0.707 for the [11] direction and 0.5 for the [01] direction \cite{ KapaklisVassilios2012Masi}. However, as shown in Fig.~\ref{Figure3}~(e), the implanted ASI exhibits higher M$_{\mathrm{rem}}$/M$_{\mathrm{sat}}$ values at temperatures below 350 K, approximately 0.8 for the [11] direction and 0.6 for the [01] direction. These values reflect the presence of a non-collinear magnetization component. 
It is important to mention that such non-collinearities in the mesospin texture have also been reported for conventional $\delta$-doped Pd(Fe) ASI lattices at low temperatures \cite{SkovdalBjörnErik2023Tewa}. However, the inhomogeneities inherent in implanted patterns seem to amplify the effect further.

Consequently, we perform micromagnetic simulations to investigate the non-collinearities in implanted mesospins. In the simulations, we model a vertex using periodic boundary conditions (see Fig.~S3), where each element consists of a core and a periphery area with distinct magnetic properties to imitate the experimental conditions (see Supplemental Material). The simulations confirm the presence of a non-collinear magnetization component at the mesospin edges for any periphery lateral width (see Supplemental Material). 
Subsequently, we calculate the ratio M$_{\mathrm{rem}}$/M$_{\mathrm{sat}}$ as a function of the angle $\theta$ between the collinear and non-collinear components, by assuming constant periphery width and non-collinearities at the whole mesospin edges (semicircles). For the calculation of the ratio M$_{\mathrm{rem}}$/M$_{\mathrm{sat}}$ the system was not allowed to relax. The results are presented in Fig.~\ref{Figure4}. For collinear magnetic texture ($\theta$ = 0), the ratio M$_{\mathrm{rem}}$/M$_{\mathrm{sat}}$ is equal to 0.707 for the [11] direction and 0.5 for the [01] direction, as expected. By introducing non-collinearities ($\theta$ > 0), M$_{\mathrm{rem}}$/M$_{\mathrm{sat}}$ increases.

\begin{figure}[t!]
\includegraphics[width=0.9\linewidth]{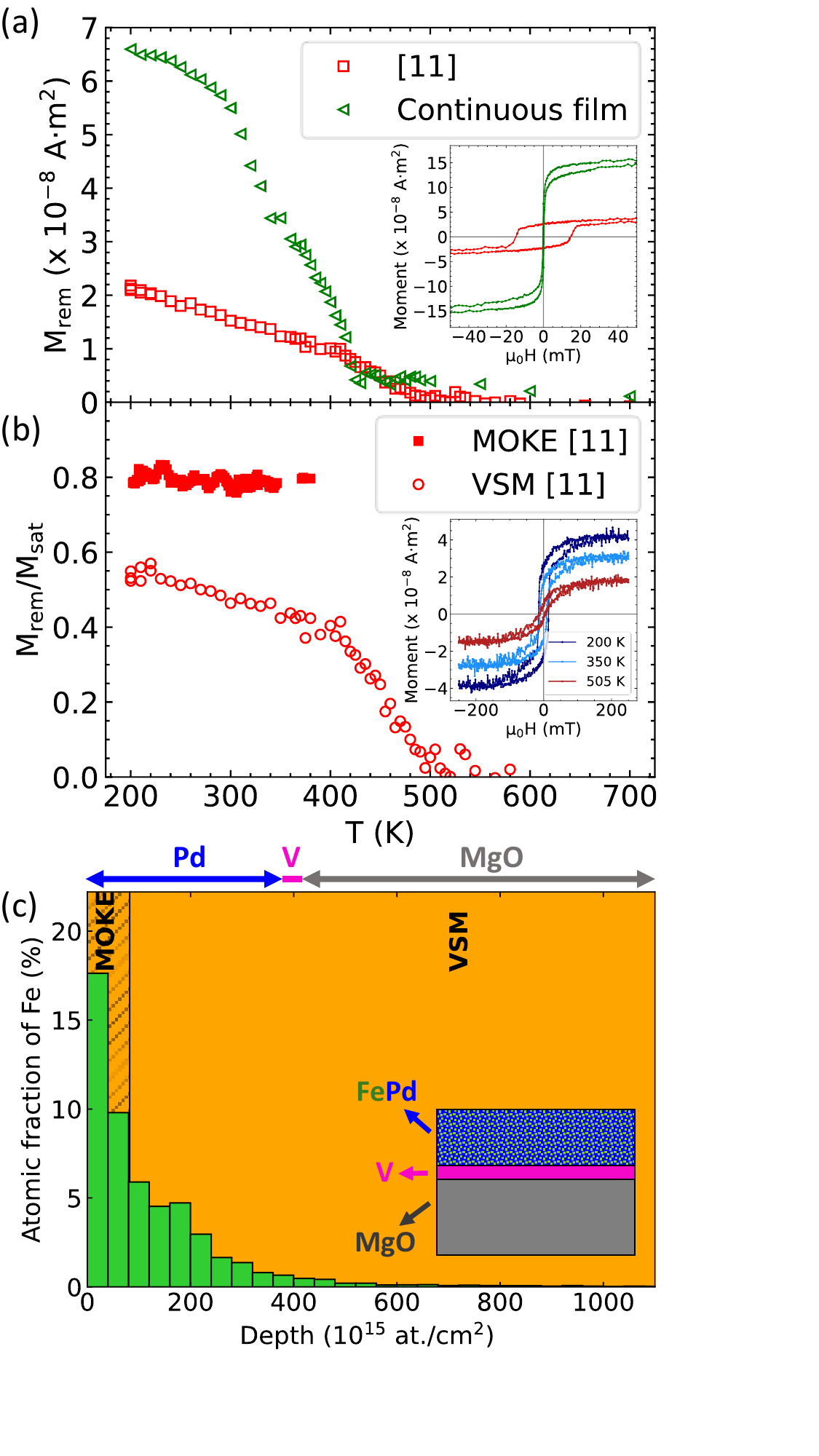}
\caption{(a) Remanent magnetization of the ASI lattice and its reference continuous film, implanted with 30~keV Fe and nominal fluence 3$\times$10$^{16}$~ions/cm$^{2}$, extracted from the VSM measurements. For both samples, the remanent magnetization of the magnet and the substrate has been corrected for. The ASI exhibits lower magnetic moment due to the smaller coverage.
(b) The ratio M$_{\mathrm{rem}}$/M$_{\mathrm{sat}}$ for the implanted ASI lattice with nominal fluence 3$\times$10$^{16}$~ions/cm$^{2}$ as a function of temperature extracted from the L-MOKE and VSM measurements. 
The saturation magnetization was determined from the hysteresis loops by averaging the values between 30~mT and 50~mT for the MOKE measurements and between 150~mT and 250~mT for the VSM measurements. 
The inset shows hysteresis loops measured by the VSM. (c) The Fe concentration depth-profile for the continuous film implanted with 30~keV Fe and nominal fluence 3$\times$10$^{16}$~ions/cm$^{2}$ measured with time-of-flight elastic recoil detection analysis (ToF-ERDA). The background shading indicates the penetration depth of the MOKE and VSM measurement systems.}
\label{Figure5}
\end{figure}

To explore if the aforementioned effects are fluence-dependent, we chose to investigate the same ASI lattice implanted with a nominal fluence of 3$\times$10$^{16}$~ions/cm$^{2}$. 
The pattern was investigated in the [11] direction and measured with the L-MOKE system at the temperature range of 200 - 400~K and the VSM system at the temperature range of 200 - 700~K. 
The remanent magnetization as a function of the temperature for the ASI lattice and its reference continuous film extracted from the VSM measurements is presented in Fig.~\ref{Figure5}~(a). The array magnetization collapses at higher temperatures compared to the film magnetization, verifying that the ordering temperature is higher for the array. 
The temperature-dependent ratio M$_{\mathrm{rem}}$/M$_{\mathrm{sat}}$ from both L-MOKE and VSM measurements is shown in Fig.~\ref{Figure5}~(b). The MOKE measurements show that M$_{\mathrm{rem}}$/M$_{\mathrm{sat}}$ has a plateau with a height of approximately 0.8, hinting towards non-collinearities in mesospin magnetic texture. 
However, the corresponding VSM measurements indicate lower values. 
As MOKE detects the surface magnetization, whereas VSM quantifies the volume magnetization, and the Fe concentration depth-profile decays from the sample surface towards the substrate [Fig.~\ref{Figure5}~(c)], the difference in M$_{\mathrm{rem}}$/M$_{\mathrm{sat}}$ indicates that the magnetic moments are more rigid close to the sample surface but fluctuate deeper. 
This observation highlights a tunable magnetic thickness with temperature, for the implanted building blocks.

In conclusion, we have demonstrated that the magnetic order in Fe-implanted artificial spin ice (ASI) lattices collapses at higher temperatures than in their reference continuous films, with the mesospins exhibiting magnetic inhomogeneities both vertically and laterally. 
The magnetic inhomogeneities along the depth enable for a tunable magnetic thickness with temperature for implanted building blocks, introducing an additional degree of freedom. This tunability may allow the mesospins to adjust their magnetic dimensionality, transitioning from 2D behavior at higher temperatures to 3D behavior at lower temperatures. In addition to that, the lateral magnetic inhomogeneities can allow modifications to the gap between the elements and their shape, enabling modifications to the dipolar coupling between building blocks after fabrication.
These findings open new pathways for tuning the static and dynamic properties of magnetic materials by leveraging fabrication processes that go beyond traditional subtractive methods such as electron-beam lithography and milling. Such advances may significantly impact future applications of magnetic metamaterials.
It was recently demonstrated that the ability to design and control magnetic textures in systems such as ASI enables the realization of tailored magnonic regimes \cite{Gartside_NatNano_2022}. These emerging textures play a crucial role in defining the accessible ground states and governing transitions between magnetization states, thereby shaping the system’s overall dynamics \cite{Sloetjes_JPCM}. The engineering of energy landscapes through fabrication techniques like the ion implantation presented here—harnessing the internal degrees of freedom in mesospins—is expected to be important in the advancement of reconfigurable magnetic metamaterials \cite{Gartside_NatNano_2022}, where dynamic behavior simultaneously functions as the readout.
Furthermore, these enhanced internal dynamics, combined with the ability to tune mesospin couplings, open promising avenues for the development of reconfigurable reservoir computing networks based on magnetic metamaterials \cite{WalczakSteven1999Hpft,Jensen2018ComputationIA, StenningKilianD.2024Noaf}. 
The thermal behavior described here is also critical for practical implementations of Ising machines using magnetic metamaterials \cite{Bhanja_NatNano_2016}. The design and precise control of their stochastic dynamics are essential prerequisites for these systems to be effectively employed in solving optimization problems \cite{MohseniNaeimeh2022Imah}.
As previously suggested \cite{Bhanja_NatNano_2016}, the readout can be achieved using spin-transfer torque magnetic tunnel junctions. An alternative approach involves performing the readout in reciprocal space. By illuminating the array with photons of suitable wavelengths, a diffraction pattern is generated from the reflected light. The positions of the diffraction peaks reveal the periodicity in real space, thereby providing access to the energy-optimized state solution.

\begin{acknowledgments}
The authors would like to thank Johan Oscarsson and Mauricio Sortica at the Uppsala Tandem Laboratory for help with ion implantations. The authors are thankful for an infrastructure grant by VR-RFI (grant number 2019-00191) supporting the accelerator operation. The authors also acknowledge support from the Swedish Research Council (projects no. 2019-03581 and 2023-06359). CV gratefully acknowledges financial support from the Colonias-Jansson Foundation, Thelin-Gertrud Foundation, Liljewalch and Sederholm Foundation. KI and UBA acknowledge funding from the Icelandic Research Fund project 2410333. We acknowledge Myfab Uppsala for providing facilities and experimental support. Myfab is funded by the Swedish Research Council (2020-00207) as a national research infrastructure. 
The PEEM-XMCD experiments were performed at CIRCE (bl24) beamline at ALBA Synchrotron with the collaboration of ALBA staff. MF acknowledges support from MICIN through grant number PID2021-122980OB-C54.
The authors are deeply grateful for the support from ReMade@ARI, funded by the European Union as part of the Horizon Europe call HORIZON-INFRA-2021-SERV-01 under grant agreement number 101058414 and co-funded by UK Research and Innovation (UKRI) under the UK government’s Horizon Europe funding guarantee (grant number 10039728) and by the Swiss State Secretariat for Education, Research and Innovation (SERI) under contract number 22.00187. VK and CV would like to thank Prof. Bj\"orgvin Hj\"orvarsson for discussions.

\end{acknowledgments}

\noindent

The data that support the findings of this study are available from the corresponding authors upon reasonable request.

%

\pagebreak
\onecolumngrid
\newpage
\begin{center}
\textbf{\large Supplemental Material: Magnetic texture control in ion-implanted metamaterials}
\end{center}
\setcounter{equation}{0}
\setcounter{figure}{0}
\setcounter{table}{0}
\setcounter{page}{1}
\makeatletter
\renewcommand{\theequation}{S\arabic{equation}}
\renewcommand{\figurename}{Supplementary FIG.}
\renewcommand{\thefigure}{{\bf S\arabic{figure}}}
\renewcommand{\bibnumfmt}[1]{[S#1]}
\renewcommand{\citenumfont}[1]{S#1}
\renewcommand{\thepage}{S-\arabic{page}}

\section{\label{Appendix_A}Laser penetration depth on Pd}
We estimate the penetration depth $d$ of a laser light with wavelength $\lambda$ = 659 nm on a Pd material. The penetration depth $d$ is the reciprocal of the absorption coefficient $\alpha$:
\begin{equation}
    d = \frac{1}{\alpha}
\end{equation}
The absorption coefficient $\alpha$ determines how far light can penetrate into a material before it is absorbed and is expressed as:
\begin{equation}
    \alpha = \frac{4 \pi k}{\lambda}
\end{equation}
where $k$ is the extinction coefficient. For Pd material and light of wavelength $\lambda$ = 659 nm, $k$ = 4.42 \cite{PolyanskiyMikhailN.2024Rdoo}. Consequently, the absorption coefficient is $\alpha$ = 0.084 nm$^{-1}$ and the penetration depth is $d$ = 11.9 nm.

\section{\label{Appendix_B}Thickness of implanted Fe}
To estimate the thickness of Fe layer implanted with nominal fluence $\Phi$ = 2.7$\times$10$^{16}$ ions/cm$^{2}$, we consider that the implanted Fe has formed a compacted layer. 
To a first approximation, the mass of a single Fe$^{+}$ ion is the same with the mass of a single Fe atom and equal to:
\begin{equation}
    m_{\mathrm{Fe}} = \frac{A_{\mathrm{Fe}}}{N_{A}} = 9.27 \times 10^{-23}\: \mathrm{g}
\end{equation}
where $A_{\mathrm{Fe}}$ is the atomic mass of Fe and $N_{A}$ is the Avogadro's number.
Consequently, the area density $\rho_{A}$, i.e total mass of Fe implanted per unit area, is:
\begin{equation}
    \rho_{A} = \Phi \times m_{\mathrm{Fe}} = 25.029 \times 10^{-7}\: \mathrm{\frac{g}{cm^{2}}}
\end{equation}
By considering that the density of Fe is $\rho_{\mathrm{Fe}}$ = 7.87 g/cm$^{3}$, the thickness of the implanted Fe as compacted layer is:
\begin{equation}
    t = \frac{\rho_{A}}{\rho_{\mathrm{Fe}}} = 3.18 \:\mathrm{nm}
\end{equation}

\section{\label{Appendix_C}Collapse of the magnetic order}
\begin{figure}[!h]
\includegraphics[width=0.8\linewidth]{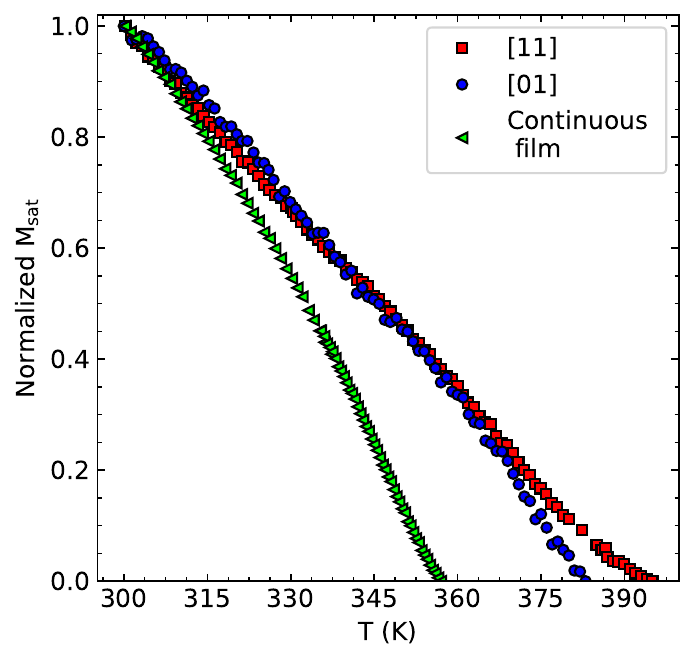}
\caption{Normalized saturation magnetization M$_{\textrm{sat}}$ as a function of temperature T for the ASI and its reference continuous film implanted with 30~keV Fe of nominal fluence 2.7$\times$10$^{16}$~ions/cm$^{2}$ measured with L-MOKE. The saturation magnetization was determined from the hysteresis loops by averaging the values between 30~mT and 50~mT.}
\label{Figure7}
\end{figure}

To verify that the ordering temperature is higher for the ASI lattice compared to its reference continuous film that was prepared and implanted next to the lattice with 30~keV Fe of nominal fluence 2.7$\times$10$^{16}$ ions/cm$^{2}$, we also examine the saturation magnetization as a function of the temperature. As shown in Fig. \ref{Figure7}, the saturation magnetization vanishes at higher temperature for the lattice compared to the continuous film, reflecting a higher ordering temperature for the ASI array.

\section{\label{Appendix_E}Vertex statistics for the remanent state [01]}
\begin{figure}[!h]
\includegraphics[width=0.7\linewidth]{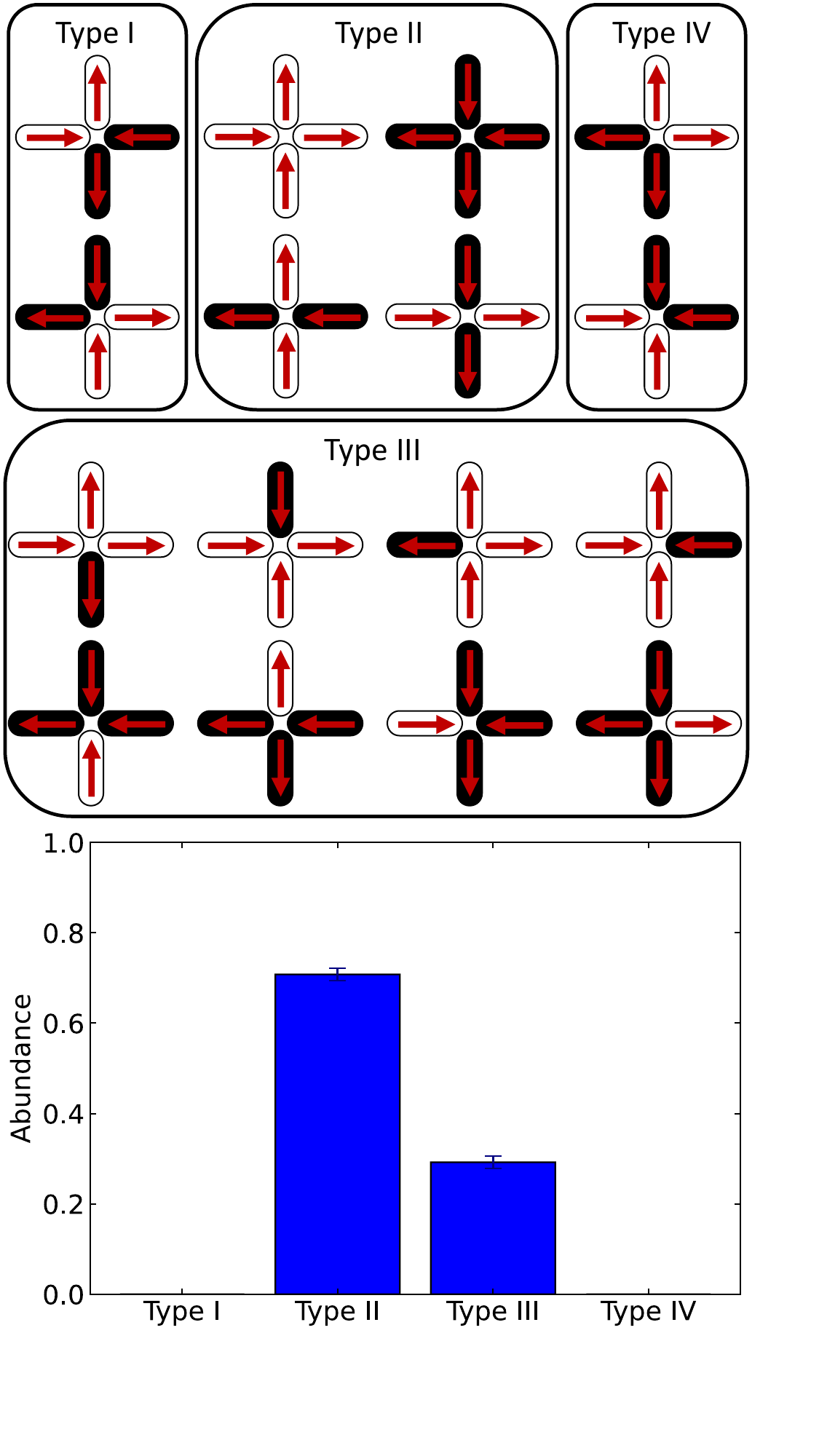}
\caption{Definition of vertex types, and the vertex statistics for the implanted ASI lattices at the remanence state after the application of a field along the [01] direction.}
\label{Figure8}
\end{figure}

To quantify the magnetic order established within the ASI lattice at the remanence state after the application of a field along the [01] direction, we determine the population of various types of vertex configuration. A vertex at which four elements converge can exhibit four distinct topologies, classified as Type I, II, III, and IV. Type I possesses the lowest energy, and Types II, III, and IV exhibit progressively higher energy. Vertex statistics are presented in Fig. \ref{Figure8}. We notice that the ASI lattice has obtained a mixture of Type II and III configurations.

\section{\label{Appendix_D}Non-collinearities at mesospin edges}
\begin{figure}[!t]
\includegraphics[width=0.7\linewidth]{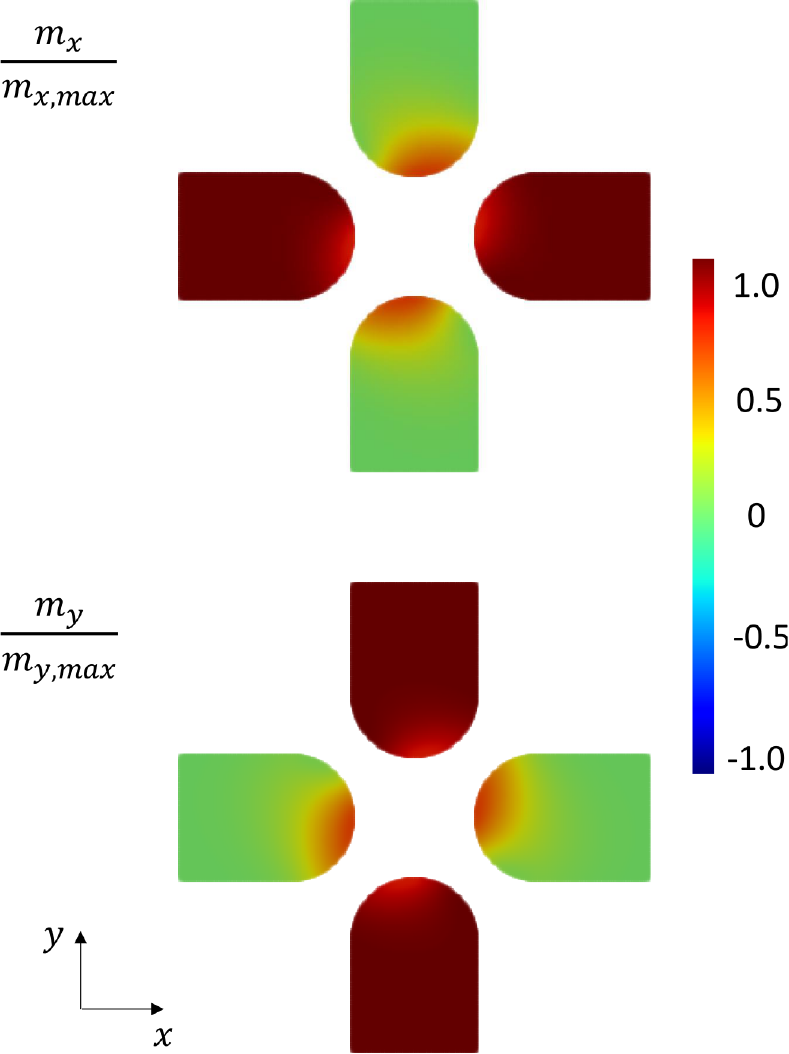}
\caption{Micromagnetic state of a vertex comprised of mesospins with inhomogeneous magnetization as shown in Fig. 2 (c). The system was allowed to relax from a state polarized in the [11] direction (Type II vertex). Normalized moment values are presented for the $x$- and $y$-direction, highlighting the sizable non-collinearities at the mesospin edges, close to the vertex center.}
\label{Figure6}
\end{figure}

We performed micromagnetic simulations to explore the non-collinearities in mesospin texture. For the simulations, we investigated a vertex using periodic boundary conditions (Fig.~\ref{Figure6}), while each element is made by a core and a periphery area.
These two areas were described by different magnetic properties, saturation magnetization and exchange stiffness constant. Specifically, the saturation magnetization and the exchange stiffness constant for the core area of the elements were chosen based on previous studies (M$_{\textrm{sat}}$~=~3.5$\times$10$^{5}$~A/m and A$_{\textrm{ex}}$ = 3.36$\times$10$^{-12}$~J/m) \cite{SkovdalBjörnErik2021Tcos}.
For the periphery, we assumed that the Fe concentration is lower by 2\% compared to the core area of the elements, and that the magnetic properties follow a linear relationship with the Fe concentration. Consequently, the saturation magnetization and the exchange stiffness constant for the periphery were considered to be equal with M$_{\textrm{sat}}$~=~3.43$\times$10$^{5}$~A/m and A$_{\textrm{ex}}$ = 3.2928$\times$10$^{-12}$~J/m respectively. The magnetization of each mesospin was assumed uniform, and subsequently the system was relaxed into its ground state. 
The simulations were performed for various periphery lateral thicknesses.
Fig. \ref{Figure6} shows a representative relaxed vertex state, confirming the presence of non-collinear magnetization component at mesospin edges.

\clearpage

\end{document}